\documentstyle[aps,preprint]{revtex}
\begin{document}
\draft
\author{Yi-shi Duan$^1$, Peng-ming Zhang$^1$\footnote{Anthor to whom
correspondence should be addressed. Email: zhpm@lzu.edu.cn} and
Hong Zhang$^2$}
\title{Topological excitation in the Fractional Quantum Hall system}
\date{\today}
\address{$^1$Institute of Theoretical Physics, Lanzhou University-Lanzhou, 730000, PRC%
\\
$^2$Department of Physics and Centre for Nonlinear Studies Hong Kong Baptist%
\\
University-Hong Kong, PRC}
\maketitle

\begin{abstract}
Two kinds of topological excitations, vortices and skyrmions, are
studied in the frame of Ginzburg-Landau theory. We obtain the
rigorous relation between the topological excitation and the order
parameters in the fractional quantum Hall systems. We also discuss
the evolution of the vortices in fractional quantum Hall systems.
\end{abstract}


\section{Introduction}

The fractional quantum Hall (FQH) effect appears in
two-dimensional (2D) election systems in a strong magnetic field.
Since its discovery in 1982 \cite{Tsui,Laughlin}, experiments on
FQH systems have continued to reveal many new phenomena and
surprises. These, together with the observed rich hierarchical
structures \cite{Haldane}, indicate that electron systems that
demonstrate a FQH effect (those systems are called FQH liquids)
contain extremely rich internal structures. There are two kinds of
topological excitations in single-layer quantum Hall systems. When
the system is fully polarized, the relevant charged quasiparticles
are topological vortices which carry the $U(1)$ topological
charges. But for a weak Zeeman coupling, it has been suggested
that \cite {Sondhi} the lowest energy quasiparticles at filling
factor $\nu =1/m\,$are skyrmions or slowly varying spin textures
which carry the $SU(2)$ topological charges.

In this paper, we consider the topological properties of vortices and
skyrmions in terms of the $\phi $-mapping topological current theory\cite
{DuanLiYang,DuanZhangLi}. The inner structure of the vortex three-current is
given with the spinless wave function in the fully polarized system.
Furthermore, the evolution of the vortex is also investigated. And one sees
that the vortices generate or annihilate at the limit points and encounter,
split, or merge at the bifurcation points of the spinless wave function.
Based on the gauge potential decomposition theory\cite{DuanU1} the skyrmion
three-current in the depolarized system is investigated from a kind of new
viewpoint.

\section{Vortex excitation}

In this section we discuss the vortex excitation of a FQH system in the
frame of GL theory. It is well-known that the topological vortices exist in
the fully polarized system, the spinless wave function can be denoted as
\begin{equation}  \label{unit1}
\psi (\vec r,t)=\phi ^1(\vec r,t)+i\phi ^2(\vec r,t)=|\phi |n(\vec r,t),
\end{equation}
where $n(\vec r,t)=n^1(\vec r,t)+in^2(\vec r,t)$ and $n^a(a=1,2)$ is the
two-dimensional unit vector field
\[
n^a(\vec r,t)=\phi ^a(\vec r,t)/|\phi |,\;\;\;\;|\phi |^2=\phi ^a\phi
^a,\;\;\;\;a=1,2.
\]
The vortex current was given by \cite{Lee1}
\[
J_v^\mu =\frac 1{2\pi }\epsilon ^{\mu \nu \lambda }\partial _\nu \overline{n}%
\frac{\partial _\lambda }in.
\]
By making use of the unit vector field $n^a$ the vortex current can be
expressed as
\[
J_v^\mu =\frac 1{2\pi }\epsilon ^{\mu \nu \lambda }\epsilon _{ab}\partial
_\nu n^a\partial _\lambda n^b,\;\;\;\;\;\;a,b=1,2.
\]
It is clear that the vortex current is identically conserved, i.e.
\begin{equation}  \label{conserv}
\partial _\mu j_v^\mu =0.
\end{equation}
By making use of the $\phi $-mapping topological current theory, this vortex
current can be rewritten in a compact form \cite{DuanLiYang,DuanZhangLi},
\begin{equation}  \label{zero1}
j_v^\mu =D^\mu (\frac \phi x)\delta (\vec \phi ),
\end{equation}
where $D^\mu (\frac \phi x)$ is the vector Jacobians of $\phi (x):$%
\begin{equation}  \label{d-v}
D^\mu (\frac \phi x)=\frac 12\varepsilon ^{\mu \nu \lambda }\varepsilon
_{ab}\partial _\nu \phi ^a\partial _\lambda \phi ^b.
\end{equation}
From the Eq. (\ref{zero1}), we can see that vortex current $j_v^\mu $ does
not vanish only at the zero points of $\phi ,$ i.e.
\begin{equation}  \label{ldfc}
\phi ^1(x^1,x^2,t)=0,\;\;\;\phi ^2(x^1,x^2,t)=0.
\end{equation}
The solutions of Eqs. (\ref{ldfc}) can be generally expressed as
\begin{equation}  \label{s1}
x^1=x_l^1(t),\;\;\;\;\;x^2=x_l^2(t),\;\;\;\;l=1,2,..,N,
\end{equation}
which represent $N$ zero points $\vec z_l(t)\;(l=1,2,...,N)$ or worldlines
of $N$ vortices in space-time. The location of $l$th vortex is determined by
the $l$th zero point $\vec z_l(t).$

According to the $\phi $-mapping topological current theory \cite
{DuanLiYang,DuanZhangLi}, one can prove that
\begin{equation}  \label{duu}
\delta ^2(\vec \psi )=\sum\limits_{l=1}^N\frac{\beta _l}{|D(\frac \phi x)|_{%
\vec z_l}}\delta ^2(\vec r-\vec z_l),
\end{equation}
where the positive integer $\beta _l$ is called the Hopf index of map $%
x->\phi .$ The meaning of $\beta _l$ is that when the point $\vec r$ covers
the neighborhood of the zero $\vec z_l$ once, the vector field $\vec \phi $
covers the corresponding region $\beta _l$ times. With the definition of
vector Jacobians (\ref{d-v}), we can obtain the general velocity of the $l$%
th vortex
\begin{equation}  \label{velocity}
v^\mu =\frac{dx_l^\mu }{dt}=\frac{D^\mu (\phi /x)}{D(\phi /x)}|_{\vec z%
_l},\;\;\;\;\;v^0=1.
\end{equation}
Then the vortex three-current $j_v^\mu $ can be written as the form of the
current and the density of the system of $N$ classical point particles with
topological charge $W_l=\beta _l\eta _l$ moving in the (2+1)-dimensional
space-time
\begin{eqnarray}
\vec{j} &=&\sum\limits_{l=1}^NW_l\vec{v}_l\delta ^2(\vec{z}-\vec{z}_l(t)),
\nonumber \\
\rho &=&j^0=\sum\limits_{l=1}^NW_l\delta ^2(\vec{z}-\vec{z}_l(t)),
\label{f564}
\end{eqnarray}
where $\eta _l=sgn(D(\phi /x)|_{\vec z_l})=\pm 1$ is the Brouwer degree $
\cite{DuanZhangLi,Hopf}.$ It is clear to see that Eq. (\ref{f564}) shows the
movement of the vortices in space-time.

\section{The generation and annihilation of vortices}

As being discussed before, the zeros of the condensate wave function $\psi $
play an important role in studying the vortices in the GL theory of the
Fractional Quantum Hall Effect. Now, we begin studying the properties of the
zero points (locations of vortices), in other words, the properties of the
solutions of Eqs. (\ref{ldfc}). As we know before, if the Jacobian
\begin{equation}  \label{dnzero}
D(\frac \phi x)=\frac{\partial (\phi ^1,\phi ^2)}{\partial (x^1,x^2)}\neq 0,
\end{equation}
we will have the isolated solutions (\ref{s1}) of Eqs. (\ref{ldfc}).
However, when the condition (\ref{dnzero}) fails, the usual implicit
function theorem is of no use. The above results (\ref{s1}) will charge in
some way and will lead to the branch process. We denote one of the zero
points as $(t^{*},\vec z_l).$ If the Jacobian
\begin{equation}  \label{n89}
D^1(\frac \phi x)|_{(t^{*},\vec z_l)}\neq 0,
\end{equation}
we can use the Jacobian $D^1(\frac \phi x)$ instead of $D(\frac \phi x)$ for
the purpose of using the implicit function theorem. Then we have an unique
solution of Eqs. (\ref{ldfc}) in the neighborhood of the points $(t^{*},\vec %
z_l)$%
\begin{equation}  \label{q4}
t=t(x^1),\;\;\;x^2=x^2(x^1),
\end{equation}
with $t^{*}=t(z_l^1).$ And we call the critical points $(t^{*},\vec z_l)$
the limit points. In the present case, it is easy to know that
\begin{equation}  \label{wqd}
\frac{dx^1}{dt}|_{(t^{*},\vec z_l)}=\frac{D^1(\phi /x)|_{(t^{*},\vec z_l)}}{%
D(\phi /x)|_{(t^{*},\vec z_l)}}=\infty
\end{equation}
i.e.
\[
\frac{dt}{dx^1}|_{(t^{*},\vec z_l)}=0.
\]
The Taylor expansion of the solution of Eq. (\ref{q4}) at the limit point $%
(t^{*},\vec z_l)$ is \cite{DuanLiYang}
\begin{equation}  \label{q5}
t-t^{*}=\frac 12\frac{d^2t}{(dx^1)^2}|_{(t^{*},\vec z_l)}(x^1-z_l^1)^2
\end{equation}
which is a parabola in the $x^1-t$ plane. From Eq. (\ref{q5}), we can obtain
two solutions $x_1^1(t)$ and $x_2^1(t),$ which give two branch solutions
(worldlines of vortices) of Eqs. (\ref{ldfc}). If $\frac{d^2t}{(dx^1)^2}%
|_{(t^{*},\vec z_l)}>0,$ we have the branch solutions for $t>t^{*},$
otherwise, we have the branch solutions for $t<t^{*}.$ These two cases are
related to the origin and annihilation of vortices.

From Eq. (\ref{wqd}), we obtain an important result that the velocity of
vortices is infinite when they are annihilating or generating, which is
gained only form the topology of the condensate wave function.

Since the topological charge of vortices is identically conserved (\ref
{conserv}), the topological charges of these two vortices must be opposite
at the limit point, i.e.,
\begin{equation}  \label{chargeI}
\beta _{l_1}\eta _{l_1}=-\beta _{l_2}\eta _{l_2},
\end{equation}
which shows that $\beta _{l_1}=\beta _{l_2}$ and $\eta _{l_1}=-\eta _{l_2}.$

\section{Bifurcation of vortex three-current}

For a limit point, it also requires $D^1(\phi /x)|_{(t^{*},\vec{z}_l)}\neq
0. $ As to a bifurcation point, it must satisfy a more complex condition at
the bifurcation point ($t^{*},\vec{z}_l$):

\begin{equation}  \label{bifa12}
\left\{
\begin{array}{c}
D(\phi /x)|_{(t^{*}, \vec z_l)}=0 \\
D^1(\phi /x)|_{(t^{*},\vec z_l)}=0
\end{array}
\right.
\end{equation}
which will lead to an important fact that the function relationship between $%
t$ and $x^1$ is not unique in the neighborhood of the bifurcation point ($%
t^{*},\vec z_l$). It is easy to see from equation
\begin{equation}  \label{bifa13}
\frac{dx^1}{dt}|_{(t^{*},\vec z_l)}=\frac{D^1(\phi /x)|_{(t^{*},\vec z_l)}}{%
D(\phi /x)|_{(t^{*},\vec z_l)}}
\end{equation}
which under the restrain (\ref{bifa12}) directly shows that the direction of
the integral curve of Eq. (\ref{bifa13}) is indefinite, i.e., the velocity
field of vortices is indefinite at the point $(t^{*},\vec z_l).$ This is why
the very point $(t^{*},\vec z_l)$ is called a bifurcation point of the
condensate wave function.

Next, we will find a simple way to search for the different directions of
all branch curves (or velocity field of vortex) at the bifurcation point.
Assume that the bifurcation point $(t^{*},\vec z_l)$ has been found from
Eqs. (\ref{ldfc}) and (\ref{bifa12}). The Taylor expansion of the solution
of Eqs. (\ref{ldfc}) in the neighborhood of the bifurcation point $(t^{*},%
\vec z_l)$ can be expressed as \cite{DuanLiYang}
\begin{equation}  \label{bifa38}
A(x^1-z_1^1)^2+2B(x^1-z_1^1)(t-t^{*})+C(t-t^{*})^2=0,
\end{equation}
which leads to
\begin{equation}  \label{bifb38}
A(\frac{dx^1}{dt})^2+2B\frac{dx^1}{dt}+C=0,
\end{equation}
and
\begin{equation}  \label{bifa39}
C(\frac{dt}{dx^1})^2+2B\frac{dt}{dx^1}+A=0,
\end{equation}
where $A,B$ and $C$ are three parameters. The solutions of Eq. (\ref{bifb38}%
) or Eq. (\ref{bifa39}) give different directions of the branch curves
(worldlines of vortices) at the bifurcation point.

The remainder component $dx^2/dt$ can be given by
\[
\frac{dx^2}{dt}=f_1^2\frac{dx^1}{dt}+f_t^2
\]
where partial derivative coefficients $f_1^2$ and $f_t^2$ have been
calculated \cite{DuanLiYang}. From these relations we find that the values
of $dx^2/dt$ at the bifurcation point $(t^{*},\vec z_l)$ are also possible
different because (\ref{bifa38}) may give different values of $dx^1/dt.$ The
above solutions reveal the evolution of vortices. Besides the encountering
of the vortices, i.e., two vortices encounter and then depart at the
bifurcation point along different branch point, it may split into several
vortices along different branch curves. On the contrary, several vortices
can merge into one vortex at the bifurcation point. The identical
conversation of the topological charge shows the sum of the topological
charge of final vortices must be equal to that of the initial vortices at
the bifurcation point, i.e.,
\[
\sum_f\beta _{j_f}\eta _{j_f}=\sum_i\beta _{j_i}\eta _{j_i},
\]
for fixed $j.$ Furthermore, from above studies we see that the generation,
annihilation and bifurcation of vortices are not gradual charges, but start
at a critical value of arguments, i.e. a sudden charge.

\section{Skyrmions in FQH effect}

The $\nu =\frac 1m$ FQH effect admits a Landau-Ginzburg description in terms
of a complex doublet of bosonic field $\Psi =(\left.
\begin{array}{c}
\psi _1 \\
\psi _2
\end{array}
\right. )$ and a statistical Chern-Simons gauge field when the FQH effect
system is depolarized \cite{Sondhi,ZhangSC,Ezawa}. We separate the magnitude
and $SU(2)$ bases of $\Psi _\sigma :\Psi _\sigma \rightarrow \rho z_\sigma $
with $\overline{z_\sigma }z_\sigma =1.$ The spin direction corresponding to $%
z_\sigma $ can be given by
\[
\Omega ^a=\overline{z_\alpha }\sigma _{\alpha _\beta }^az_\beta
,\;\;\;\;a=1,2,3
\]
where $\sigma ^a$ are the Pauli matrices. The skyrmion three-current is \cite
{Lee1}
\[
j_s^\mu =\frac 1{4\pi }\epsilon ^{\mu \nu \lambda }\epsilon _{abc}\Omega
^a\partial _\nu \Omega ^b\partial _\lambda \Omega ^c,\;\;\;\;\;\;\mu ,\nu
,\lambda =1,2,3.
\]
It is well-known that this current can be expressed as \cite{Fradkin}
\begin{equation}  \label{sky5}
j_s^\mu =\frac 1{4\pi }\epsilon ^{\mu \nu \lambda }(\partial _\nu A_\lambda
-\partial _\lambda A_\nu )
\end{equation}
where $A_\mu =\frac i2(z_\alpha ^{*}\partial _\mu z_\alpha -z_\alpha
\partial _\mu z_\alpha ^{*}),$ is essentially the Berry connection \cite
{Stone} with
\[
D_\mu \zeta =\partial _\mu \zeta -iA_\mu \zeta ,
\]
in which $\zeta =\zeta ^1+i\zeta ^2$ is a complex scalar field defined by
the connection $A_\mu $. As one shown in \cite{DuanU1}, the $U(1)$
connection can be decomposed by the complex scalar field $\zeta $ as
\[
A_\mu =\epsilon ^{ab}\partial _\mu m^am^b+\partial _\mu \theta ,\;\;\;a=1,2
\]
in which $\theta $ is only a phase factor and $m^a=\zeta ^a/\sqrt{\zeta
^b\zeta ^b}$ are the unit vector. With this connection the current (\ref
{sky5}) can be obtained
\begin{equation}  \label{sky6}
j_s^\mu =\frac 1{2\pi }\epsilon ^{\mu \nu \lambda }\epsilon _{ab}\partial
_\nu m^a\partial _\lambda m^b.
\end{equation}
Obviously, the current (\ref{sky6}) is conserved. Following the $\phi $%
-mapping theory, it can be rigorously proved that
\[
j_s^\mu =\delta ^2(\vec \zeta )D^\mu (\frac \zeta x),
\]
which is just similar to the vortex current $j_v^\mu $. But we have to point
that the vector field $\zeta (x)$ in the skyrmions current is different from
the spinless wave function $\phi (x)$ in the vortex current.

When the vector field $\vec \zeta $ possesses $l$ zeros, denoted as $%
z_i(i=1,2,...,l),$ the skyrmions current can also be expressed by the
winding number $W_i=\beta _i\eta _i$%
\[
j_s^\mu =\sum_{i=1}^lW_i\delta ^2(\vec x-\vec z_i)\frac{dx^\mu }{dt}|_{z_i},
\]
which gives the density of skyrmions charge
\[
\rho =\sum_{i=1}^lW_i\delta ^2(\vec x-\vec z_i).
\]
Furthermore, the total charge of the system can be rewritten as
\[
Q=\int \rho (x)d^2x=\sum_{i=1}^lW_i.
\]
It is obvious that there exist $l$ isolated skyrmions of which the
$i$th skyrmion possesses charge $W_i.$ And the skyrmion
corresponds to $\eta _i=+1, $ while the anti-skyrmion corresponds
to $\eta _i=-1.$

\section{Conclusion}

First, in terms of the spinless wave function of the full
polarized system, the inner topological structure of the vortex
excitations is discussed with the $\phi $-mapping topological
current theory. It is shown that the vortices in the FQH system
are generated from $\phi=0$ and their topological charges are
quantized under the condition $D(\phi /x)\neq 0$, i.e. the
vortices emerge in a "natural way" as a solution of a topological
current theory, instead of being put "by hand", as usually done in
more phenomenological theories. One also shows that the charge of
the vortex is determined by Hopf indices and Brouwer degrees.
Second, we have studied the evolution of the vortices and
concluded that there exist crucial cases of branch processes in
the evolution of the vortices when $D(\phi /x)=0,$ i.e., $\eta _l$
is indefinite. This means that the vortices are generated or
annihilated at the limit points and are encountered, split, or
merged at the bifurcation points of the wave function of the full
polarized system, which shows that the vortices system is unstable
at the branch points. Meantime, these conditions give simple rules
to consider the nonlinear behavior in all sort of similar system.
Third, with the skyrmions three-current given in \cite{Lee1} we
obtain the inner structure of skyrmion excitations by making use
of the decomposition of gauge potential theory. Finally, we would
like to point that all the results in this paper have been
obtained only from the viewpoint of topology without using any
particular hypothesis. They will help those experts to find a
branch point and the concrete branch process in the neighborhood
of it, and give a deep insight into fractional quantum Hall
system.

\section*{Acknowledgment}

This work was supported by the National Natural Science Foundation of China
and the Doctoral Foundation of the People's Republic of China.

\end{document}